\documentclass[aps,prb,amsmath,amssymb,twocolumn,letterpaper,%
         footinbib,bibnotes,showpacs,reprint,balancelastpage,raggedbottom,shortbibliography,%
         citeautoscript,floatfix]{revtex4-1}

\usepackage[usenames,dvipsnames,table]{xcolor}
\usepackage[pdftex,
  breaklinks,             %
  bookmarks = false, %
  pdfpagemode   = UseNone,%
  pdfstartview  = FitH,     %
  pdfstartpage  = 1,      %
  colorlinks    = true,   %
]{hyperref}
\hypersetup{
    linkcolor=RubineRed,          
    citecolor=ForestGreen,        
    filecolor=Mulberry,      
    urlcolor=RoyalBlue           
}
\usepackage{amssymb}
\usepackage{graphicx}
\usepackage{float}
\usepackage{enumerate}
\usepackage{microtype}
\usepackage{xspace}


\begin{document}

\author{Liang-Feng Huang}
  \affiliation{Department of Materials Science and Engineering, Northwestern University, Evanston, IL 60208, USA}
\author{Xue-Zeng Lu}
  \affiliation{Department of Materials Science and Engineering, Northwestern University, Evanston, IL 60208, USA}
\author{Emrys Tennessen}
  \affiliation{Department of Materials Science and Engineering, Northwestern University, Evanston, IL 60208, USA}
\author{James M.\ Rondinelli}
\email{jrondinelli@northwestern.edu}
  \affiliation{Department of Materials Science and Engineering, Northwestern University, Evanston, IL 60208, USA}

\title{An efficient ab-initio quasiharmonic approach for the thermodynamics of solids}

\begin{abstract}
A first-principles approach called the {\it{self-consistent
quasiharmonic approximation}} (SC-QHA) method is formulated to
calculate the thermal expansion, thermomechanics, and thermodynamic
functions of solids at finite temperatures with both high efficiency
and accuracy. 
The SC-QHA method requires fewer phonon calculations than the conventional QHA method, and also 
facilitates the convenient analysis of the microscopic origins of macroscopic 
thermal phenomena. 
The superior performance of the SC-QHA method is systematically examined by
comparing it with the conventional QHA method and experimental measurements on  
silicon, diamond, and alumina. 
It is then used to study the effects of pressure on the anharmonic lattice 
properties of diamond and alumina.
The thermal expansion and thermomechanics of Ca$_3$Ti$_2$O$_7$, which is a recently 
discovered important ferroelectric ceramic with a complex crystal structure that is computationally
challenging for the conventional QHA method, are also calculated using the formulated 
SC-QHA method. 
The SC-QHA method can significantly reduce the
computational expense for various quasiharmonic thermal properties especially when there are a large
number of structures to consider or when the solid is structurally 
complex. 
It is anticipated that the algorithm will be useful for a variety of fields, 
including oxidation, corrosion, high-pressure physics, ferroelectrics, and high-throughput structure
screening when temperature effects are required to accurately describe realistic properties.
\end{abstract}


\maketitle


\section{Introduction}
Accurately simulating various anharmonic properties, i.e.,
thermal expansion and thermomechanics, of solids is important for obtaining a 
deep understanding of their plentiful thermal behaviors and for
their realistic applications. 
The anharmonic properties can be derived from the volume and temperature dependences 
of the phonon spectra calculated using density-functional theory (DFT) \cite{Martin2004}. 
The most popular approach is the quasiharmonic
approximation method (QHA) \cite{Baroni2010,Togo2010,Togo2015}, 
where only the volume 
dependence is considered for the phonon anharmonicity, and temperature is assumed to indirectly affect phonon vibrational 
frequencies through thermal expansion.
%
Here, the phonon spectra of about ten or more volumes are usually required for
a typical QHA simulation, and the thermal expansion and
thermomechanics are derived by fitting the free energy-volume
relationship. 
In some cases, e.g., at high
temperatures, high-order anharmonicity caused by 
multi-phonon coupling cannot be omitted as in the QHA method, and some
more complicated and time-consuming methods, e.g.,
molecular dynamics
\cite{Alfe2001,Ackland2002,Vocadlo2002,Grabowski2009,Grabowski2011,Hellman2011,Kong2012,Hellman2013},
self-consistent ab-initio lattice dynamics,
\cite{Souvatzis2008,Souvatzis2009} perturbative/nonperturbative renormalized
harmonic approximations, \cite{Rousseau2010,Errea2011,Errea2013,Michel2015} and vibrational
self-consistent field calculations \cite{Monserrat2013}, can be used to obtain the
temperature-dependent phonons. Nonetheless, approximately ten or more
volumes of such phonon spectra are also required to accurately calculate the
thermal expansion and thermomechanics with the high-order anharmonicities.
Phonon calculations based on DFT forces are always time consuming, and
prior to the actual calculation, various computational parameters \cite{Martin2004,Hafner2008} also
need to be carefully tested to ensure convergence of the vibrational frequencies and anharmonicity,
including the pseudopotentials, cutoff
energy, $k$-mesh density, energy and force convergence thresholds,
and supercell size in the small-displacement method
\cite{Kresse1995,Parlinski1997} or the $q$-mesh density in the
density-functional perturbation theory approach 
\cite{Baroni2001,Giannozzi2009}. 
The general rule-of-thumb requiring ten or more
volumes will make the anharmonic simulation, even when utilizing the simplest
QHA method, rather computationally expensive, especially in some
condensed matter fields where a large number of structures must be considered 
or the compound under study has a large unit cell, low symmetry, and numerous  inequivalent atoms: 
\begin{enumerate}[(i)]
\item In the fields of solid oxidation
and corrosion, there are always many 
compounds (elements, oxides, hydroxides, oxyhydroxides, etc.) to consider 
\cite{Beverskog1996,Beverskog1997_1,Beverskog1997_2,Beverskog1997_3,Beverskog1997_4,Chivot2008}
and each composition may have many polymorphs
\cite{ICSD,Jain2013,Saal2013};
\item In the high-pressure physics field, not only a wide range of volumes
but also a large number of complex phases should be examined \cite{Oganov2006,Wang_Ma_2010,Wang2015};
\item For the metallic alloys field, the thermodynamics and mechanics of many phases at 
variable composition and temperature are always of concern \cite{Maisel2012,Li2012,Yuan2014};
\item In the perovskite oxides  \cite{Rondinelli2012,Benedek2015,Young2015}, ternary ceramics exhibit 
complex structures and large unit cells. The phonon calculations for
an individual structure is already quite time consuming, not to mention the
calculation of anharmonic properties in low-symmetry polymorphs; and 
\item In high-throughput screening and  materials design 
\cite{Curtarolo2012,Curtarolo2013,Rondinelli2015,Rondinelli2015_2}
when including  temperature effects, a huge number of compositions
and structures should be calculated with a high efficiency-to-accuracy
ratio.
\end{enumerate}
To this end, these diverse fields require an efficient
method to accelerate the investigation of the anharmonic
properties of related solids at finite temperatures.
In this work, we formulate an ab-initio method, called the {\it{self-consistent
quasiharmonic approximation}} (SC-QHA) method, for achieving 
fast anharmonic calculations with high accuracy within the quasiharmonic approximation. 
Only the phonon spectra of two or three
volumes are required in a SC-QHA calculation,
which usually is much faster than the conventional 
QHA method. 
We carefully test the SC-QHA method using 
prototypical silicon, diamond, and alumina, and then also study the pressure
effect on the anharmonic properties of diamond and alumina. 
Finally, we apply the SC-QHA method to accurately calculate the thermal expansion and thermomechanics 
of the structurally complex hybrid-improper ferroelectric 
Ca$_3$Ti$_2$O$_7$. 
Apart from the high efficiency, we show that the SC-QHA method is also very convenient for deciphering the 
microscopic physical origins of lattice dynamical and thermodynamic phenomena.
Moreover, it can be readily transferred beyond the
quasiharmonic realm to speed up the accurate
first-principles simulation of thermal effects for the benefit of multiple 
fields in condensed-matter physics.

\section{Thermodynamics and Computation}\label{methods}
\subsection{Theoretical Basis}
The total Gibbs free energy ($G_{tot}=F_{tot}+PV$) of a crystal unit
cell is expressed as
\begin{equation}{\label{Equ_G_tot}}
G_{tot}(P,T)=F_e(V,T)+F_{ph}(V,T)+PV\,,
\end{equation}
where $P$, $V=V(P,T)$, and $T$ are the external pressure, unit-cell volume, and temperature, respectively; 
$F_e$ and $F_{ph}$ are the electronic and phononic Helmholtz
free energies, respectively.
To conveniently present
the basic algorithm and efficiency of the SC-QHA method, only
nonmagnetic insulators are considered here, where the electronic
excitation and magnetic excitations are neglected. 
The transferability of the SC-QHA algorithm for solids with more complex degrees of 
freedom are discussed below.
Therefore, $F_e(V,T)$ here equals the electronic energy $E_e(V)$ calculated from
density functional theory (DFT) and $F_{ph}(V,T)$ is expressed as
\begin{equation}{\label{Equ_F_ph}}
F_{ph}=\frac{1}{N_q}\sum_{q,\sigma}
\left\{\frac{\hbar\omega_{q,\sigma}}{2}+k_BT\log{\left[1-\exp\left(-\frac{\hbar\omega_{q,\sigma}}{k_BT}\right)\right]}\right\}\,,
\end{equation}
where $k_B$ is the Boltzmann constant, $N_q$ is the number of considered reciprocal points,
$\omega_{q,\sigma}$ is the vibrational frequency of the
$\sigma$-th phonon branch at the reciprocal coordinate $q$. 

The equilibrium state under a specified external pressure $P$ fulfills
the relationship 
\begin{equation}{\label{Equ_dG_0}}
\left.\frac{{\delta}G_{tot}}{{\delta}V}\right|_{P,T}=0.
\end{equation}
Combining Eqs.\ \ref{Equ_G_tot}, \ref{Equ_F_ph} and
\ref{Equ_dG_0}, we obtain an expression for the unit-cell volume
\begin{equation}{\label{Equ_V_T}}
V(P,T)=\left[\frac{dE_e(V)}{dV}+P\right]^{-1}\times\frac{1}{N_q}\sum_{q,\sigma}U_{q,\sigma}\gamma_{q,\sigma},
\end{equation}
where $U_{q,\sigma}$ and $\gamma_{q,\sigma} =-\frac{V}{\omega_{q,\sigma}}\frac{d\omega_{q,\sigma}}{dV}$ are
the internal energy and Gr\"uneisen parameter of the phonon mode
($q$,$\sigma$). 
(To guarantee that $\gamma$ is calculated from the
phonon modes with the same symmetry, $k{\cdot}p$ theory is used
to identify and match the phonon branches obtained from different volumes according to the similarity of each mode's eigenvector.\cite{Huang2014})
The physics underlying Eq. \ref{Equ_V_T} is due to 
the balance between the external pressure $P$ and internal pressure,
i.e., electronic pressure ($P_e=-\frac{dE_e}{dV}$) plus the 
anharmonic phonon pressure
($P_\gamma=-\frac{dF_{ph}}{dV}=\frac{1}{VN_q}\sum_{q,\sigma}U_{q,\sigma}\gamma_{q,\sigma}$), such that 
\begin{equation}{\label{Equ_P_Pe_Pgamma}}
P=P_e(V)+P_\gamma(V,T).
\end{equation}

In the quasiharmonic approximation,
\cite{Baroni2010,Togo2010,Togo2015} $\omega$ only 
depends on
$V$ such that the $\omega$--$V$ relationship can be
described by a Taylor expansion (up to second order) as 
\begin{equation}{\label{Equ_Omega_V}}
\omega(V)=\omega(V_0)+\left(\frac{d\omega}{dV}\right)_0\Delta{V}+\frac{1}{2}\left(\frac{d^2\omega}{dV^2}\right)_0\Delta{V}^2,
\end{equation}
where $V_0$ is the reference volume and $\Delta{V}=V-V_0$. Then, we
can derive the volume dependence of $\gamma$ as 
\begin{equation}{\label{Equ_gamma_V}}
\gamma(V)=-\frac{V}{\omega}\left[\left(\frac{d\omega}{dV}\right)_0+\left(\frac{d^2\omega}{dV^2}\right)_0\Delta{V}\right].
\end{equation}
The calculation of the $n$-th order derivative of $\omega$
(i.e., ${d^n\omega}/{dV^n}$) requires the phonon spectra of $n+1$
volumes. 
With Eqs.\ \ref{Equ_V_T}, \ref{Equ_Omega_V}, and
\ref{Equ_gamma_V}, the temperature-dependent unit-cell
volume can be obtained in a self-consistent manner and it is this 
formalism which we call herein the self-consistent quasiharmonic approximation (SC-QHA)
method.
The complete derivation steps for the SC-QHA method, as well as the formula in 
the next section, can be found in the online Supplemental Material. 
%

%
The SC-QHA method can also be viewed as an improved nonlinear Gr\"uneisen model that is implemented in a self-consistent way. 
In a conventional Gr\"uneisen model \cite{Wallace1998,Xie1999,Mounet2005,Barrera2005}, the linear $\omega$--$V$ relationship
usually is considered, and during the calculation of the thermal-expansion coefficient $\alpha=\frac{1}{V}\frac{dV}{dT}$,  the parameters 
$V$, $\omega$, and $\gamma$ are treated as constant. In addition, the zero-point vibration contribution to 
$V$ is also absent in the Gr\"uneisen model for $\alpha(T)$. 
Although a similar nonlinear $\omega$--$V$ relationship to that given 
shown in Eq.\ \ref{Equ_Omega_V} has been used by Debernardi et al.\ for $\alpha(T)$ before\cite{Debernardi2001}, 
the contribution of zero-point vibrations to $V$ were omitted. 
Herein, both the zero-point vibrational contribution and nonlinear $\omega$--$V$ relationship are treated in 
the SC-QHA method (by Eqs.\ \ref{Equ_V_T} and \ref{Equ_Omega_V}).   

%
In principle, when we have an analytical expression for $\omega(V)$ (Eq.\ \ref{Equ_Omega_V}),  $V(P,T)$ could be 
directly derived by minimizing the analytical $G_{tot}$ (Eq.\ \ref{Equ_G_tot})
with respect to $V$ prior to 
using Eq.\ \ref{Equ_V_T}. 
%
This approach should have a comparable numerical efficiency as the self-consistent method. 
However, the SC-QHA method is very convenient analytics method for which considerable physical information (as present in Eq.\ \ref{Equ_V_T}) 
can be obtained through the accompanying the iteration process or from the converged results; the SC-QHAD allows 
us to understand related properties and to further design materials.
Indeed, the lowest-order SC-QHA method with $\gamma$ kept constant, has been successfully
used to reveal various anharmonic mechanisms in two-dimensional materials \cite{Huang2014,Huang2013,Huang2014_2,Huang2015,Huang2015_2}, 
while we will show in this work that the higher-order SC-QHA method is necessary for three-dimensional solids at high temperatures
(e.g., $\gtrsim1000$ K).

\subsection{Computational Algorithm}

\begin{figure}
\scalebox{1.}[1.]{\includegraphics{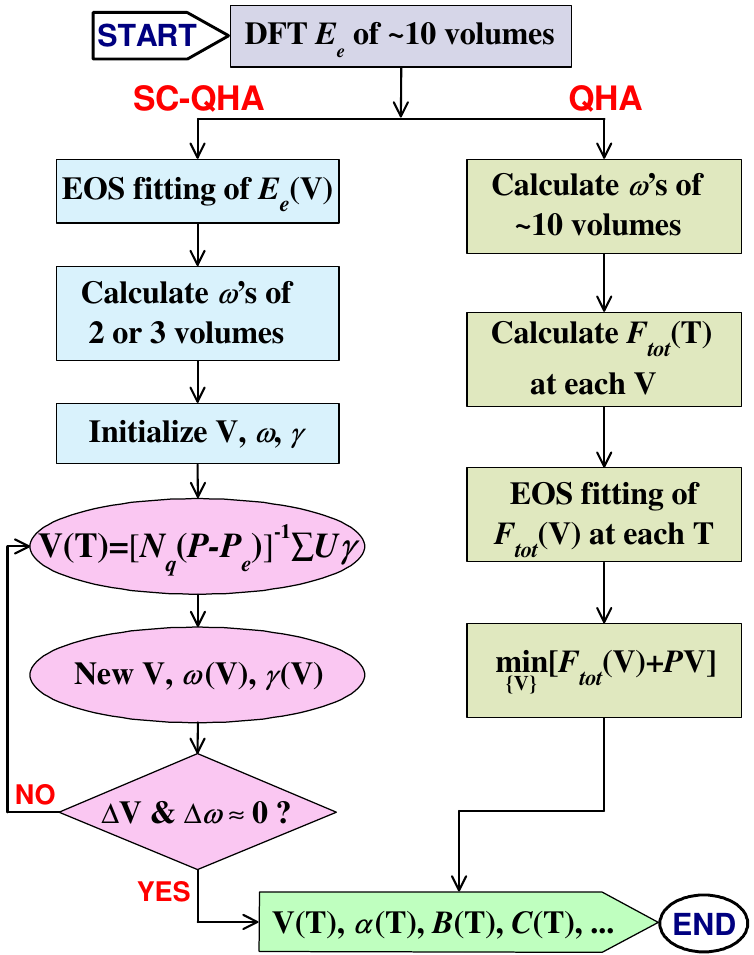}}
\caption{\label{Fig_work_flow} The workflow chart for the SC-QHA method (left) compared to the 
conventional QHA method (right). Details of their algorithms are given in the main text.}
\end{figure}

The computational protocol for the SC-QHA method is depicted in Fig.\ \ref{Fig_work_flow}, and 
described below:
\begin{enumerate}[(1)]
\item First, only the electronic energies ($E_e$) for approximately ten or more volumes are 
calculated using DFT. This step is required for both the SC-QHA and the conventional QHA
methods.
\item Next, the calculated $E_e$'s are fit by an
equation of state (EOS). There are various forms available,
i.e., Birch-Murnaghan equation, modified Birch-Murnaghan equation,
Birch equation, logarithmic equation, Murnaghan equation, Vinet
equation, and Morse equation \cite{Shang2010_3}. The fourth-order
Birch-Murnaghan EOS is used here to produce an analytical $E_e(V)$ 
and $\frac{dE_e}{dV}$ expression. 
\item Phonon spectra of two and three volumes 
are calculated 
to obtain $(\frac{d\omega}{dV})_0$ and $(\frac{d^2\omega}{dV^2})_0$, respectively. 
The 1st-SC-QHA method only requires $(\frac{d\omega}{dV})_0$, whereas the 2nd-SC-QHA requires 
both $(\frac{d\omega}{dV})_0$ and $(\frac{d^2\omega}{dV^2})_0$.
\item Prior to the self-consistent cycle, reasonable initial values for the unit-cell
volume, phonon frequencies, and Gr\"uneisen parameters are specified. 
To have a nonzero $\frac{dE_e}{dV}$ in the
denominator of Eq. \ref{Equ_V_T}, the initial $V$ is set to be
the 0.2\%-expanded DFT equilibrium $V$, which is used to calculate
the initial vibrational frequencies ($\omega$, Eq. \ref{Equ_Omega_V}) and Gr\"uneisen parameters ($\gamma$, Eq. \ref{Equ_gamma_V}).
\item The initial $\omega$ and $\gamma$ values are used to calculate a
new $V$ using Eq.\ \ref{Equ_V_T}, and the new $V$ is taken to 
recalculate phonon frequencies and Gr\"uneisen parameters using Eq.\ \ref{Equ_Omega_V} and \ref{Equ_gamma_V}, respectively. 
Then, the new $\omega$'s and $\gamma$'s are used
to further update the volume $V$, and this kind of self-consistent
iteration continues until $V$ and the $\omega$'s are
converged to within an acceptable threshold (e.g., $10^{-4}\%$). 
\item The $V(T)$ relationship is then obtained when the specified temperature window
is scanned during the self-consistent calculation, where the calculated
$V$, $\omega$'s, and $\gamma$'s at one temperature are used
as the initial values for the next temperature. 
\end{enumerate}

After obtaining converged $V$, $\omega$'s, and $\gamma$'s, then other
thermodynamic properties may be calculated, e.g., the 
thermal-expansion coefficient, bulk modulus, heat capacity, free
energy, entropy, etc.
From Eq. \ref{Equ_V_T}, the volume thermal-expansion
coefficient ($\alpha=\frac{1}{V}\frac{dV}{dT}$) is derived to be
\begin{equation}{\label{Equ_alpha}}
\alpha=\frac{1}{N_qVB_T}\sum_{q,\sigma}C_{q,\sigma}\gamma_{q,\sigma}\,,
\end{equation}
where $B_T$ is the isothermal bulk modulus, and $C_{q,\sigma}$ is the 
isovolume heat capacity of the $(q,\sigma)$ phonon mode. $B_T$ can be expressed
as the summation of four components
\begin{equation}{\label{Equ_B_T}}
B_T=B_e+B_\gamma+B_{\Delta\gamma}+P_\gamma\,,
\end{equation}
where $B_e=V\frac{d^2E_e}{dV^2}$ is the electronic bulk
modulus; $P_\gamma=P-P_e$, as described above, is the
anharmonic-phonon pressure, a first-order component; $B_\gamma$ and $B_{\Delta\gamma}$ are
the second-order components, and
$B_{\Delta\gamma}$ results from the change of $\gamma$. $B_\gamma$
and $B_{\Delta\gamma}$ are expressed as
\begin{eqnarray}{\label{Equ_Bgamma}}
B_\gamma &=& -\frac{1}{N_q}\sum_{q,\sigma}{\frac{dU_{q,\sigma}}{dV}\gamma_{q,\sigma}}   \\\nonumber
&=& \frac{1}{N_qV}\sum_{q,\sigma}\left(U_{q,\sigma}-C_{q,\sigma}T\right)\gamma_{q,\sigma}^2
\end{eqnarray}
and
\begin{eqnarray}{\label{Equ_Bgamma}}
B_{\Delta\gamma} &=& -\frac{1}{N_q}\sum_{q,\sigma}{U_{q,\sigma}\frac{d\gamma_{q,\sigma}}{dV}}   \\\nonumber
 &=& -\frac{1}{N_qV}\sum_{q,\sigma}U_{q,\sigma}\left[(1+\gamma_{q,\sigma})\gamma_{q,\sigma}-\frac{V^2}{\omega_{q,\sigma}}\frac{d^2\omega_{q,\sigma}}{dV^2}\right],
\end{eqnarray}
where $\frac{d^2\omega_{q,\sigma}}{dV^2}$ equals zero in the 1st-SC-QHA
method. In addition, by obtaining the thermal expansion and
isothermal thermomechanics, the isoentropic thermomechanics can also be readily derived \cite{Wang2010_2,Nye1985,Barrera2005}. Note that the
mechanical properties obtained from experimental measurements always
reside between the isothermal and isoentropic limits, and this
work mainly focuses on the introduction of the SC-QHA method rather than 
focusing on the detailed mechanical relationships.
%

%
To realize a faster SC-QHA calculation, we self-consistently solve Eq.\ \ref{Equ_V_T} for $V$ at very low temperature (e.g., 0.1 K),
and then use $\alpha(T)$ to obtain $V$ at higher temperatures, i.e., 
\begin{equation}{\label{Equ_alpha_V}}
V(T+{\Delta}T)=(1+\alpha\cdot{\Delta}T)V(T),
\end{equation}
which help us avoid the self-consistent calculation for each temperature. A temperature increment (${\Delta}T$) of $\lesssim2.0$ K 
can yield converged results.
%

%
The numerical implementation of the code is written in FORTRAN and 
can be accessed via a GitHub Repository at \href{http://github.com/MTD-group}{\sffamily http://github.com/MTD-group} under GPLv3.

\subsubsection{Comparison to the Conventional QHA Approach}
In the conventional QHA method (see Fig.\ \ref{Fig_work_flow}, right column),
\cite{Baroni2010,Togo2010,Togo2015} the phonon spectra for approximately ten
or more volumes, especially for accurate thermomechanics, are required to be calculated, from which the $F_{tot}$ of
each volume within a specified temperature window is calculated.
Then, an EOS is used to fit the calculated $F_{tot}(V)$ at each
considered temperature. Finally, $G_{tot}=F_{tot}+PV$ is minimized with
respect to $V$ at each considered temperature, which yields various thermodynamic properties,
e.g., $V(T)$, $\alpha(T)$, and $B(T)$. 
%
Such process is needed in both testing convergence parameters and in performing meaningful QHA calculations.
However, the 1st-SC-QHA method using phonon spectra of only two volumes can be efficiently used 
in such test calculations. 
After the optimal numerical and computational parameters are determined, 
the 2nd-SC-QHA method using phonon spectra of three volumes may then be used to simulate accurate thermodynamic quantities.
%

\subsubsection{Extension to Metallic and Magnetic Crystals}

Herein nonmagnetic insulators are considered,
however, in metals \cite{Wang2004,Grabowski2007} and magnetic solids
\cite{Kormann2008,Wang2009,Wang2010,Shang2010,Shang2010_2,Alling2010,Alling2010_2,Wang2011,Liu2011,Kormann2012,Steneteg2012,Kormann2014}, electronic and magnetic excitations, as well as spin-phonon coupling, at finite temperatures should be
considered in computing the thermodynamic properties. 
For the SC-QHA simulation of a metal, we need to additionally treat the
derivatives of $F_e(V,T)$ in a similar way as $F_{ph}(V,T)$;  the
electronic Gr\"uneisen parameters \cite{Barrera2005} also need to be calculated. 
The main difference is that the Fermi-Dirac distribution
should be used for the electrons in $F_e$, while the Bose-Einstein distribution is used for the phonons in $F_{ph}$.

For magnetic solids, the ensemble of
magnetic configurations and the spin-phonon coupling should be correctly considered at finite temperatures,
while the volume dependence of the phonon spectra is similarly calculated as that for nonmagnetic solids. 

\subsubsection{Extension to High-Order Anharmonic Cases}

%
In the quasiharmonic approximation, where $\omega=\omega(V)$, only the low-order anharmonicity of 
phonons is considered \cite{Walle2002}. When multi-phonon interactions are considerable, especially at very high temperatures, 
the phonon frequency has an additional explicit temperature dependence, i.e., $\omega=\omega(V,T)$, and 
some expensive methods beyond the quasiharmonic approximation, as described in the Introduction, 
should be used to capture such high-order anharmonicity. 
In the calculation of high-order anharmonic thermal expansion and thermomechanics, 
if the free-energy EOS fitting algorithm in the conventional QHA method is adopted, 
the anharmonic phonon spectra of ten or more volumes are required for an accurate fit.
Alternatively, the Taylor-expansion and self consistency algorithms in the SC-QHA method
can be used to efficiently describe the volume dependence of the anharmonic phonons, and the computational expense 
for high-order anharmonic cases will be significantly reduced.
%

\subsection{Density-Functional Calculations}
The Vienna \emph{Ab Initio} Software Package (VASP) \cite{Hafner2008,Kresse1996,Kresse1996_2} 
is used to calculate the energies, forces, and stresses at the DFT level. 
The projector augmented-wave (PAW) method \cite{Blochl1994,Kresse1999} is used to
treat the interaction among the core and valence electrons and a planewave cutoff energy of 
800 eV is also used. 
The electronic exchange-correlation potential is described by the PBEsol
\cite{Perdew2008,Perdew2009} parameterization, which we find 
generally gives more accurate thermodynamic quantities.\cite{Huang_Rondinelli_2015} 
We use a reciprocal-grid density of
$\approx\frac{30}{a}\times\frac{30}{b}\times\frac{30}{c}$, where $a$, $b$, 
and $c$ are the lattice constants of the periodic cell scaled by the unit of angstrom. The
convergence thresholds for the electronic energy and force are
$5\times10^{-8}$ eV and $5\times10^{-4}$ eV\,\AA$^{-1}$, respectively. The
atomic positions, cell volume, and cell shape are globally optimized
for the equilibrium states, and only the atomic positions and cell
shape are optimized for a solid at a compressed/expanded volume.

\begin{figure}[t]
\scalebox{1.0}[1.0]{\includegraphics{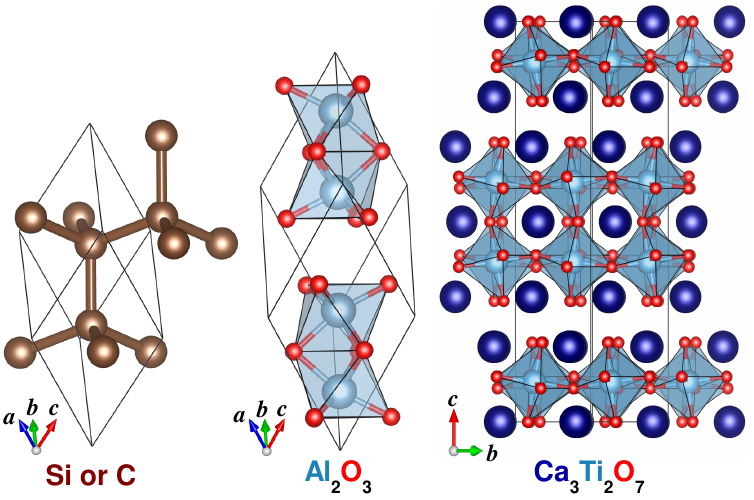}}
\caption{\label{Fig_Structures} The structures for silicon (diamond),
alumina, and Ca$_3$Ti$_2$O$_7$, where the unit cells are indicated by the black empty parallelepipeds.}
\end{figure}

The PHONOPY code \cite{Togo2010,Togo2015} is used to calculate phonon spectra,
where the small-displacement method \cite{Kresse1995,Parlinski1997} is
implemented, and the atomic displacement amount is set to be 0.03
\AA. The supercells for the phonon calculation of Si, C,
Al$_2$O$_3$, and Ca$_3$Ti$_2$O$_7$ are $5\times5\times5$ (250
atoms), $3\times3\times3$ (54 atoms), $2\times2\times2$ (80 atoms),
and $2\times2\times1$ (192 atoms) times their respective unit
cells (Fig. \ref{Fig_Structures}, see also Supplemental Material). 
For the conventional QHA
calculation, 14, 11, and 18 volumes are considered for Si, C, and
Al$_2$O$_3$ (see Supplemental Material). 

The unexpected large supercell of Si is required for its
accurate anharmonic properties according to our fast 1st-SC-QHA benchmark calculations; 
the underlying physical origin is due to the existence of variable negative Gr\"uneisen parameters in its Brillouin zone
(see Supplemental Material). 
The LO--TO splitting has negligible effect on the anharmonic properties according to our 
SC-QHA test calculations  on Al$_2$O$_3$ (Supplemental Material), because only the LO
because only the LO phonons in a very small
reciprocal space close to the $\Gamma$ point are affected
\cite{Shang2007,Shang2010_4}. 

For the 2nd-SC-QHA calculation, the phonon
spectra of the equilibrium, 3.0\%-compressed, and 3.0\%-expanded
volumes are used for Si, C, and Al$_2$O$_3$, where the later two
volumes are used for the 1st-SC-QHA calculation. 
When
Ca$_3$Ti$_2$O$_7$ is under expansion, portions of the acoustic phonon branches 
will be spuriously softened and become dynamically unstable in the DFT
calculation (Supplemental Material). However, in
experimental measurement \cite{Senn2015}, Ca$_3$Ti$_2$O$_7$ has no
phase instability during its thermal expansion up to temperatures as high as its
decomposition temperature (1150\,K). 
Thus, it should be the increased phonon-phonon coupling upon heating that 
stabilizes those acoustic phonons and makes them behave normally, which is quite similar to 
what occurs in TiO$_2$ \cite{Montanari2004,Shojaee2010,Mitev2010}. To avoid this
kind of phonon anomaly in expanded Ca$_3$Ti$_2$O$_7$, the phonon
spectra of the equilibrium, 1.5\%-compressed, and 3.0\%-compressed
volumes are used in the SC-QHA calculation of Ca$_3$Ti$_2$O$_7$. 
In addition, the conventional QHA method cannot be used when there is any 
instability in the quasiharmonic DFT phonons. This is also a numerical
advantage of the SC-QHA method compared to the conventional QHA method. 

\begin{figure*}
\scalebox{0.75}[0.75]{\includegraphics{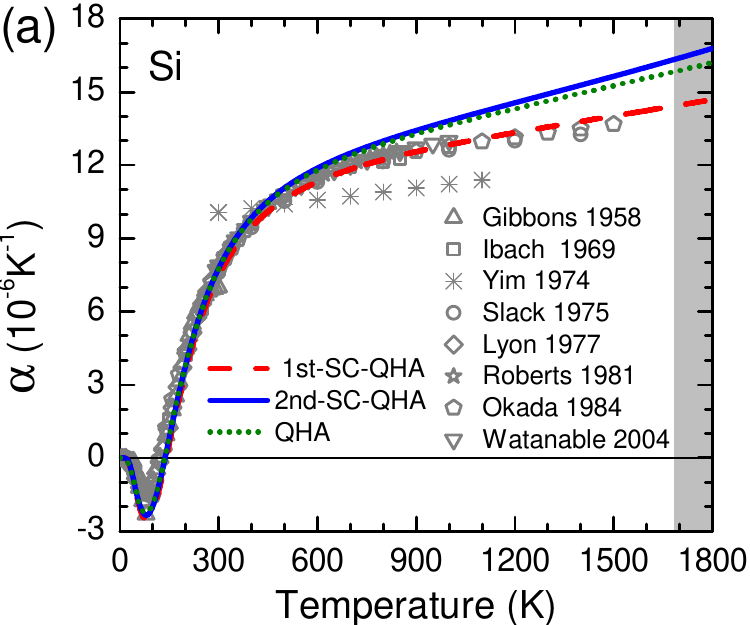}}
\scalebox{0.75}[0.75]{\includegraphics{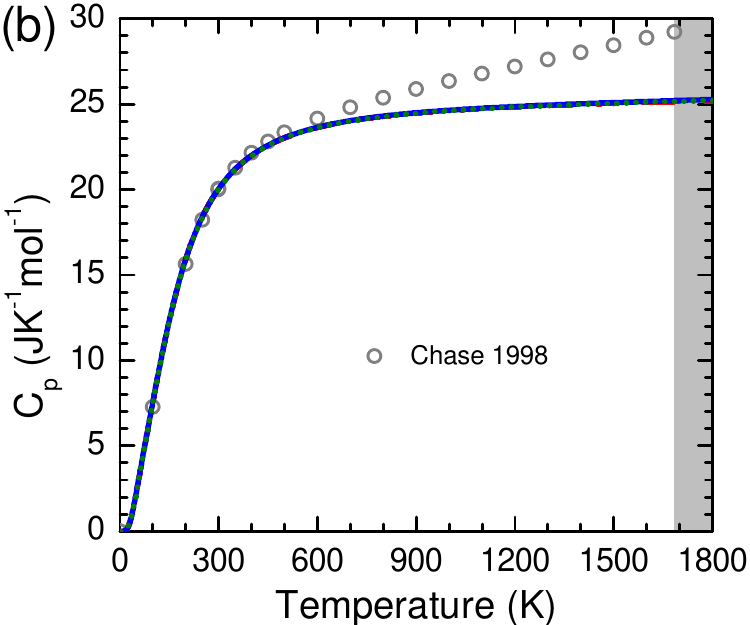}}
\scalebox{0.75}[0.75]{\includegraphics{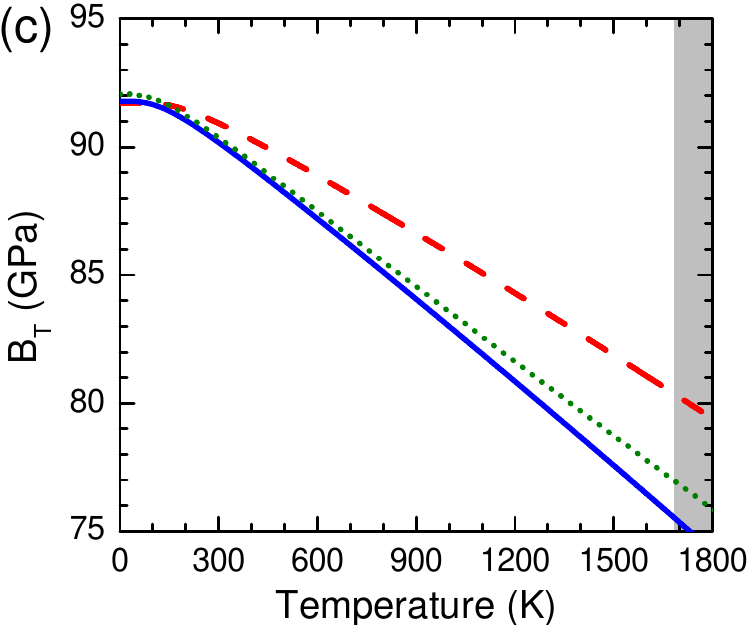}}\\
\scalebox{0.75}[0.75]{\includegraphics{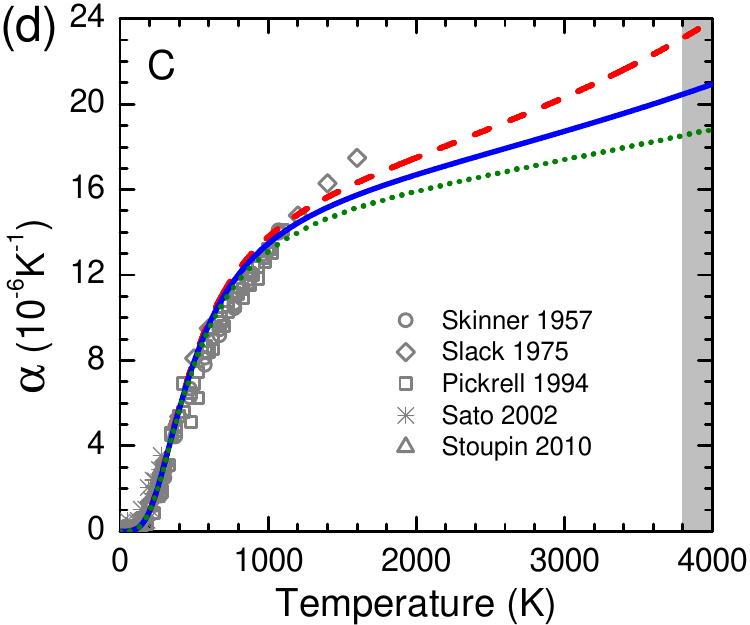}}
\scalebox{0.75}[0.75]{\includegraphics{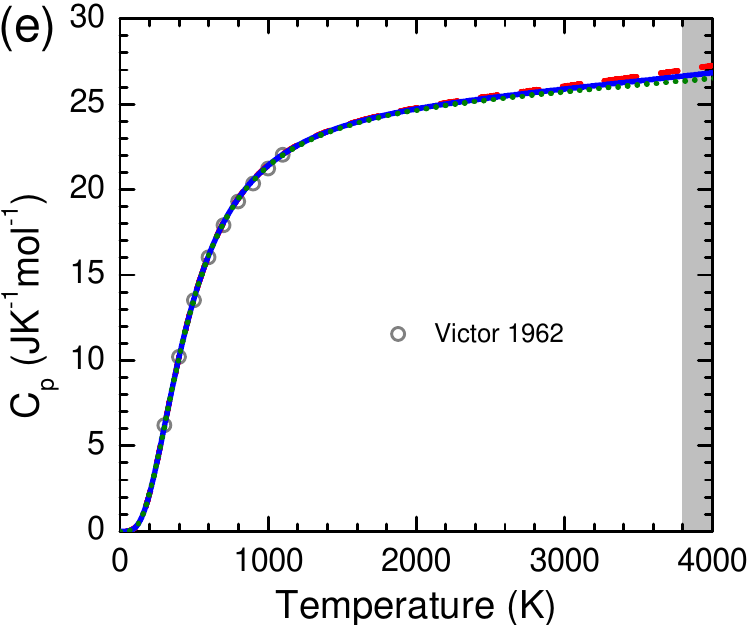}}
\scalebox{0.75}[0.75]{\includegraphics{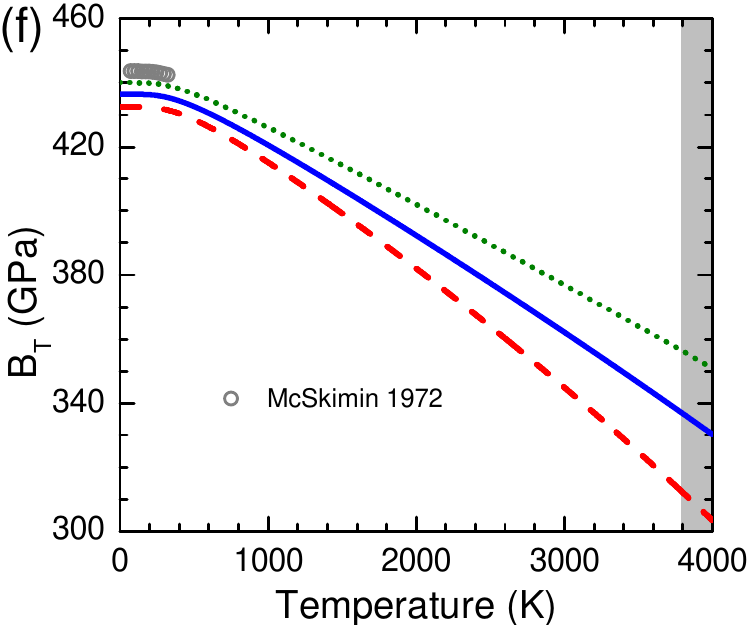}}\\
\scalebox{0.75}[0.75]{\includegraphics{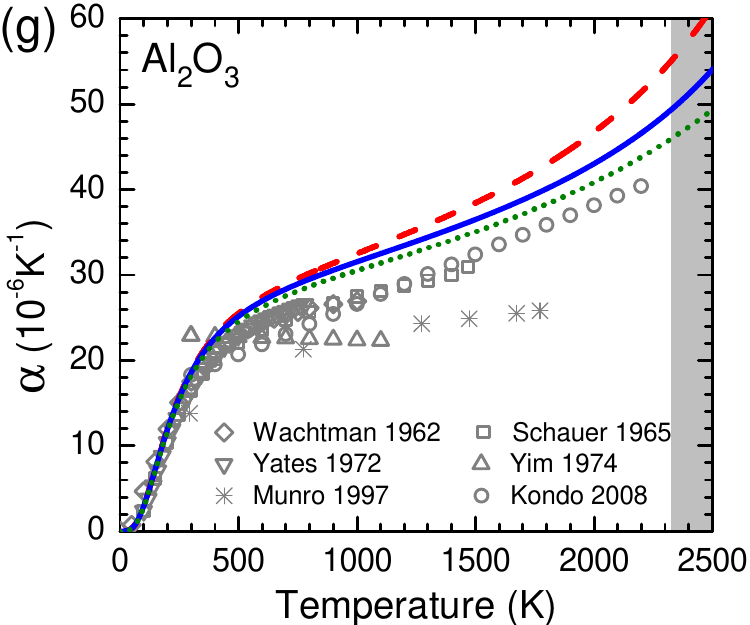}}
\scalebox{0.75}[0.75]{\includegraphics{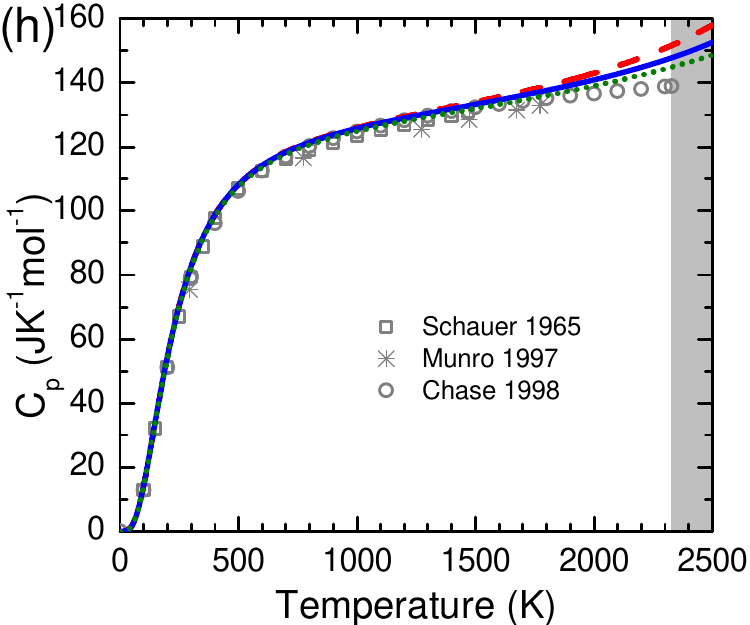}}
\scalebox{0.75}[0.75]{\includegraphics{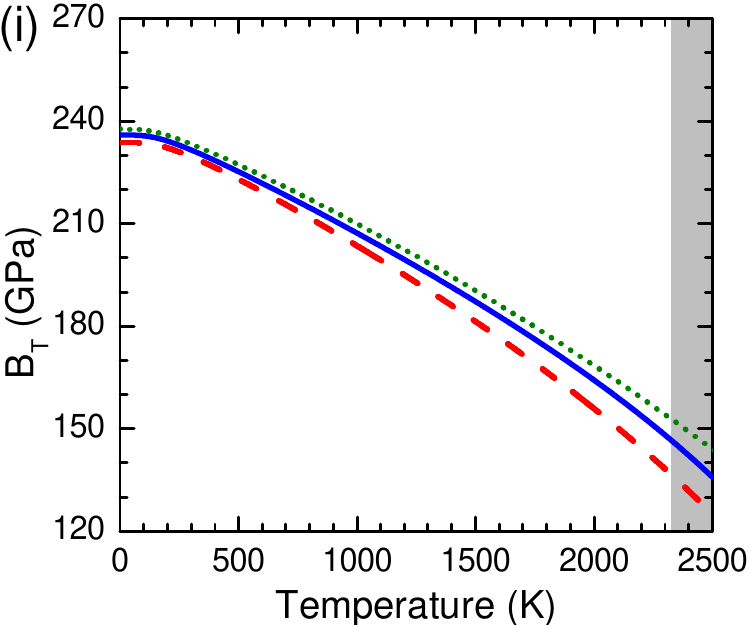}}
\caption{\label{Fig_Si_C_Al2O3_results} The volume thermal-expansion
coefficient ($\alpha$), isobaric heat capacity ($C_p$), and
isothermal bulk modulus ($B_T$) of (a--c) Si, (d--f) diamond, and
(g--i) alumina, where the areas with temperatures above the melting points are shaded in gray. The experimental,
\cite{Skinner1957,Gibbons1958,Victor1962,Wachtman1962,Schauer1965,Ibach1969,McSkimin1972,Yates1972,Yim1974,Slack1975,Lyon1977,Roberts1981,Okada1984,Pickrell1994,Munro1997,Chase1998,Sato2002,Watanable2004,Kondo2008,Stoupin2010}
1st-SC-QHA, 2nd-SC-QHA, and QHA results are presented by symbols,
dashed red lines, solid blue lines, and dotted green lines, respectively. The experimental results are also 
tabulated for accessibility in the Supplemental Materials. 
}
\end{figure*}

\begin{figure*}
\scalebox{0.75}[0.75]{\includegraphics{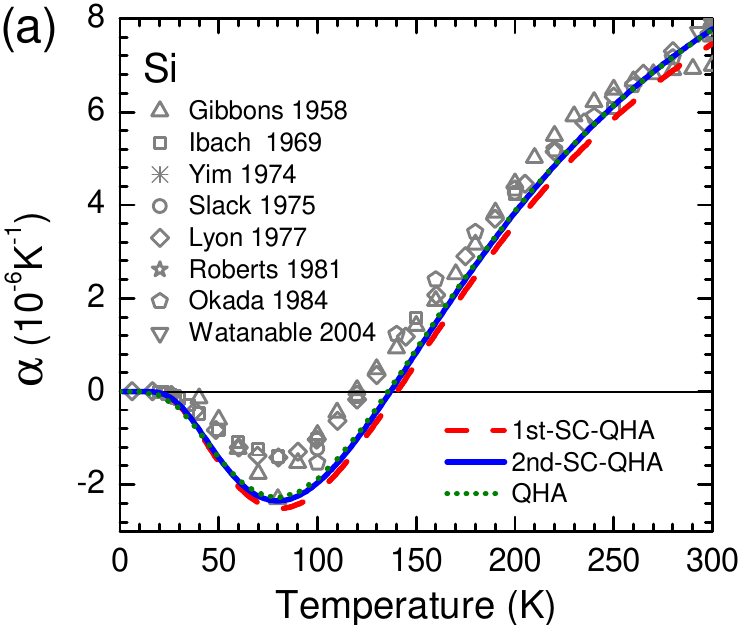}}
\scalebox{0.75}[0.75]{\includegraphics{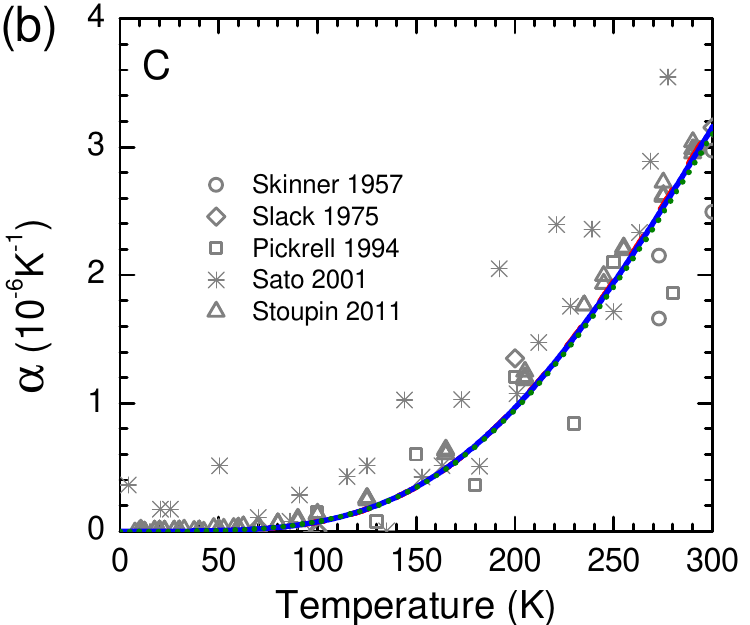}}
\scalebox{0.75}[0.75]{\includegraphics{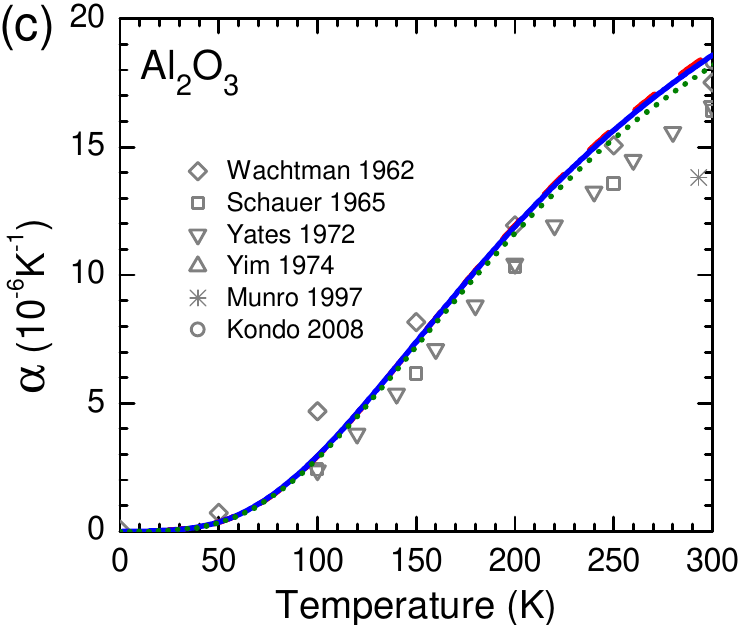}}\\
\scalebox{0.75}[0.75]{\includegraphics{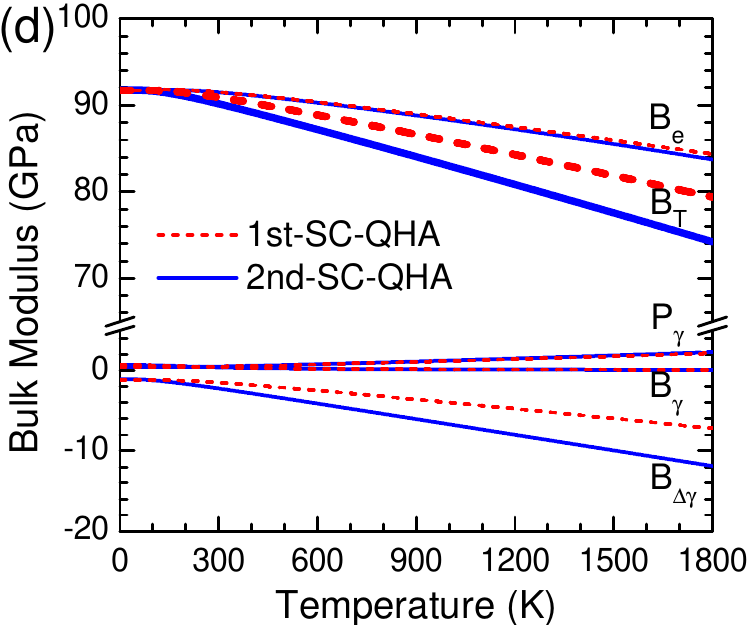}}
\scalebox{0.75}[0.75]{\includegraphics{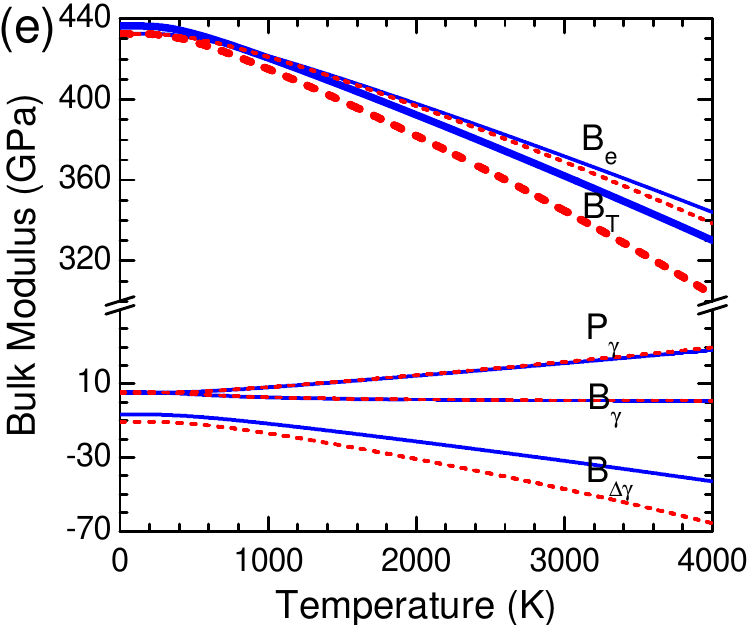}}
\scalebox{0.75}[0.75]{\includegraphics{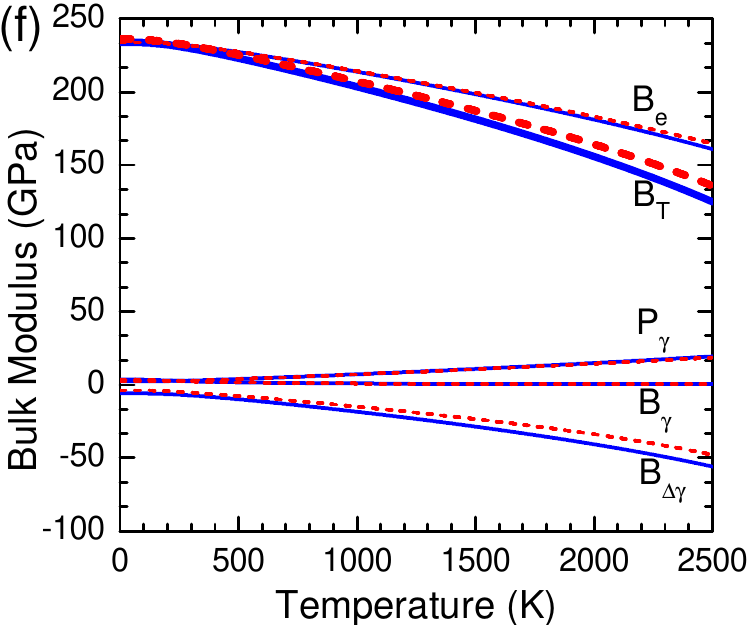}}
\caption{\label{Fig_Si_C_Al2O3_details} The variations of $\alpha$
at low temperatures and $B_T$ components ($B_e$, $B_\gamma$,
$B_{\Delta\gamma}$, and $P_\gamma$) for (a,d) silicon, (b,e)
diamond, and (c,f) alumina. In panels (a--c), the three calculated
curves nearly overlap and are close to the experimental results 
\cite{Skinner1957,Gibbons1958,Victor1962,Wachtman1962,Schauer1965,Ibach1969,McSkimin1972,Yates1972,Yim1974,Slack1975,Lyon1977,Roberts1981,Okada1984,Pickrell1994,Munro1997,Chase1998,Sato2002,Watanable2004,Kondo2008,Stoupin2010}.
}
\end{figure*}

\section{Results and discussion}
\subsection{Algorithm Benchmarks}
The calculated volume thermal-expansion coefficient ($\alpha$),
isobaric heat capacity ($C_p=C_v+TV_TB_T\alpha^2$) \cite{Nye1985}, and isothermal bulk modulus ($B_T$)
obtained from the 1st-SC-QHA, 2nd-SC-QHA, and QHA methods for
silicon (Si), diamond (C), and alumina (Al$_2$O$_3$) under zero
pressure and at temperatures from 0 K to the melting point, $T_m$, are shown
in Fig.\ \ref{Fig_Si_C_Al2O3_results}. 
%
The low-temperature variations in $\alpha$($T$) and detailed analysis of $B_T$ are given 
in Fig.\ \ref{Fig_Si_C_Al2O3_details}.
The available experimental results for Si, C, and Al$_2$O$_3$
\cite{Skinner1957,Gibbons1958,Victor1962,Wachtman1962,Schauer1965,Ibach1969,McSkimin1972,Yates1972,Yim1974,Slack1975,Lyon1977,Roberts1981,Okada1984,Pickrell1994,Munro1997,Chase1998,Sato2002,Watanable2004,Giles2005,Kondo2008,Stoupin2010}
are also collected to compare with the calculated values in Fig.\ 
\ref{Fig_Si_C_Al2O3_results} and \ref{Fig_Si_C_Al2O3_details}. 
The raw data is available in digital format in the Supplemental Material. 

\subsubsection{Thermal Expansion Coefficients}
All three methods generally perform well for $\alpha$ when compared
with experimental measurements [Fig.\ \ref{Fig_Si_C_Al2O3_results}(a,d,g)].
If we use the QHA results as the theoretical reference for our proposed method, then 
we find that the 2nd-SC-QHA approach has a higher accuracy than the 1st-SC-QHA
method. 
At temperatures of $\leqslant300$\,K, the $\alpha$($V$) curves
from these three methods are in excellent agreement with each other and the
experimental results, allowing for a scattering of
$\lesssim2\times10^{-6}$ K$^{-1}$ in experimental data [Fig.
\ref{Fig_Si_C_Al2O3_details}(a,b,c)]. 
At temperatures of $\leqslant\frac{1}{4}$
$T_m$, the 2nd-SC-QHA (1st-SC-QHA) thermal expansion coefficients deviate from the QHA
ones only by 0.4\% (-3.6\%), 2.7\% (4.9\%), and 2.6\% (4.3\%) for
Si, C, and Al$_2$O$_3$, respectively [Fig.
\ref{Fig_Si_C_Al2O3_results}(a,d,g)]. 
Thus, both the 1st-SC-QHA and
2nd-SC-QHA methods have high accuracy for the thermal expansion at
relatively low temperatures. 
At higher temperatures, e.g.,
$\frac{2}{3}T_m$, the thermal expansion coefficients obtained from 
the 2nd-SC-QHA (1st-SC-QHA) deviate from
the QHA ones by 1.5\% (-6.5\%), 6.3\% (13.0\%), and 4.2\% (10.1\%)
for Si, C, and Al$_2$O$_3$, respectively [Fig.
\ref{Fig_Si_C_Al2O3_results}(a,d,g)], which indicates that
the 2nd-SC-QHA method should be preferred to the 1st-SC-QHA method when seeking to achieve 
an accurate simulation of the thermal expansion at relatively high
temperatures. 

Here we note that it is also quite challenging to measure $\alpha$ accurately
in experiment, especially at high temperatures, and the scatter 
in the experimental data always significantly increases with
temperature. 
In fact, the theory-experiment discrepancy is even smaller than
the experimental uncertainty [Fig.
\ref{Fig_Si_C_Al2O3_results}(a,d,g)]. Apart from some possible
theoretical factors (e.g., the exchange-correlation potential 
and the quasiharmonic approximation itself) causing the inaccuracies in the simulated thermal expansion coefficients,
there are also many experimental factors that can influence the measurement
accuracy.
%
First, thermal expansion always varies with the crystalline orientation
\cite{Ibach1969,Munro1997}. When the crystalline
orientation is not well characterized, the derived thermal expansion
may vary among single crystals, powders, and polycrystals.
Second, the measured samples are readily contaminated by
various impurities (e.g., Cu, Fe, Mg, Ca, etc.), which
may have profound effect on thermal expansion\cite{Thewlis1955,Ibach1969,Yim1974,Pickrell1994,Reeber1996,Munro1997,Sato2002}.
%
Next, the precipitation of some metastable phases in the sample at relative high temperatures may
influence the thermal expansion. For example, there are many
polymorphs (e.g., $\alpha$, $\theta$, $\gamma$, and $\delta$
phases) of Al$_2$O$_3$ \cite{Levin1998,Andersson2005}, and the
relative stability among the polymorphs 
correlates with temperature and impurity concentration
\cite{Andersson2005,Huang2015_3}. It also has been found that the
metastable $\theta$-Al$_2$O$_3$ has a smaller thermal expansion than
the stable $\alpha$-Al$_2$O$_3$ \cite{Shang2010_4}, which may be
related with the fact that the experimental thermal expansion is always
smaller than the simulated one for $\alpha$-Al$_2$O$_3$ [Fig.
\ref{Fig_Si_C_Al2O3_results}(g)].
%
%

Moreover there are many methods to measure the thermal expansion of
solids, e.g., push-rod dilatometer \cite{Roberts1981},
capacitance dilatometer \cite{Ibach1969,Lyon1977,Falzone1982},
interferometric dilatometer
\cite{Gibbons1958,Ibach1969,Roberts1981,Watanable2004}, and X-ray
diffraction
\cite{Skinner1957,Yim1974,Slack1975,Okada1984,Pickrell1994,Sato2002,Giles2005,Stoupin2010}.
Although an accuracy from $10^{-6}$ to $10^{-8}$ K$^{-1}$ may be
declared for each measurement, the discrepancy between 
different measurements still can be on the order of $10^{-6}$
K$^{-1}$.
Last, the thermal-expansion coefficient is derived from the
lattice-constant variation with temperature, where 
analytical functions are used for the data fitting
\cite{Yim1974,Lyon1977}. When the data is insufficiently large 
and exhibits obvious scatter, the derived $\alpha$ will 
also depend on the chosen fitting function. For example, a difference of several
percentages can be readily found when refitting the data reported by
Yim \cite{Yim1974}.
%

%

Most of the available experimental thermal expansion coefficients for these compounds 
are reported in earlier reports (e.g., 1950--1980s), and high-temperature data
are still scarce, which makes stringent experiment--experiment and
theory--experiment comparisons difficult. Therefore, accurate
experimental measurements (especially for high temperatures) on
high-quality samples are still needed to precisely understand the
thermal expansion of many solids. Nonetheless, according to our current
comparison, we find that relatively high accuracy is achieved by the 
2nd-SC-QHA method (or even by the 1st-SC-QHA method), which only
requires the phonon spectra of three (or two) volumes, rather than
ten or more volumes as in the conventional QHA method.

\subsubsection{Isobaric Heat Capacity}

In addition to the good performance of the SC-QHA method for the thermal 
expansion coefficients, both the 1st-SC-QHA and
2nd-SC-QHA methods also yield highly accurate isobaric heat capacities for Si, C,
and Al$_2$O$_3$, with respect to the QHA results [Fig.\ 
\ref{Fig_Si_C_Al2O3_results}(b,e,h)]. 
The simulated $C_p$ values are also consistently agree with the 
experimental ones for C and Al$_2$O$_3$
[Fig. \ref{Fig_Si_C_Al2O3_results}(e,h)], although the experimental
$C_p$ for Si gradually deviates from the simulated one at temperatures $\gtrsim
500$ K, e.g., by 8.3\% at $\frac{2}{3}T_m$ (1124 K). 
%

Usually the experiment--theory discrepancy for Si is  ascribed 
to the omission of high-order anharmonicity beyond the quasiharmonic approximation, and 
an {\it{ad-hoc}} anharmonic correction of 8\% for the heat capacity has been suggested \cite{Trivedi1985,Kagaya1993}. 
However, such a  simple anharmonic correction for heat capacity will adversely influence the accuracy of the theoretical $\alpha$: 
When the quasiharmonic heat capacity is corrected by 8\%, the quasiharmonic $\alpha$ also correspondingly needs to
be increased by about 8\% (Eq.\ \ref{Equ_alpha}) \cite{Kagaya1987}. Because the quasiharmonic $\alpha$ for Si is already higher than the experimental one. e.g., by 8\% at $\frac{2}{3}T_m$, Fig.\ \ref{Fig_Si_C_Al2O3_results}(a), such a correction will further increase the theory-experimental discrepancy in $\alpha$ by $\sim$16\% at $\frac{2}{3}T_m$.
%

%
These results then motivate for a more systematic and improved 
understanding of the anharmonic properties of Si. 
Apart from obtaining perhaps more precise experimental measurements on high-quality samples, 
state-of-art ab-initio simulations of the high-order anharmonicity are needed; here, 
our SC-QHA algorithm may be used to speed up the calculation of the volume dependence of the anharmonic phonon 
spectra needed for those higher level theories.

\subsubsection{Isothermal Bulk Modulus}

An anticipated thermal softening of materials is observed in the temperature dependence of 
$B_T$ [Fig.\ \ref{Fig_Si_C_Al2O3_results}(c,f,i)]. 
From 0 K to $\frac{2}{3}T_m$, the softening of the isothermal bulk moduli for Si, C, and 
Al$_2$O$_3$ are 6.8, 69.6, and 55.3 GPa from the 1st-SC-QHA method, 
10.1, 59.9, and 51.3 GPa from the 2nd-SC-QHA method, 
and 9.7, 51.3, and 49.7 GPa from the QHA method, respectively. 
Although each method results in the same qualitative chemical trends for $B_T$, the 
2nd-SC-QHA method gives a higher quantitative accuracy than the 1st-SC-QHA method 
with respect to the QHA method. 

We understand this improved precision obtained by the 2nd-SC-QHA method as follows: 
The $B_T$ is calculated from the second derivative of the free energy; thus, it is sensitive to both 
the method type and  the number and range of cell volumes chosen in the QHA method. 
Nonetheless, the difference between the 2nd-SC-QHA and QHA results is already smaller than
the numerical uncertainty in the QHA method (Supplemental Material). 

In addition, the SC-QHA method enables us to decompose $B_T$ into four contributing
components (Eq.\ \ref{Equ_B_T}), i.e., $B_e$, $P_\gamma$, $B_\gamma$, and $B_{\Delta\gamma}$, where $B_{\Delta\gamma}$
results from the change of the Gr\"uneisen parameter. 
The bulk moduli for Si, C, and Al$_2$O$_3$, as well as 
their corresponding components, are shown in
Fig.\ \ref{Fig_Si_C_Al2O3_details}(d,e,f), where the 1st-SC-QHA and the 
2nd-SC-QHA results are compared. 
It can be clearly discerned that the difference between the bulk moduli obtained from the 
1st-SC-QHA and the 2nd-SC-QHA mainly originates from deviations in the calculated $B_{\Delta\gamma}$, 
indicating the importance of the nonlinear vibrational frequency variation in solid thermomechanics. 
Moreover, $B_e$ is consistently larger than $B_T$, indicating that
the excitation of anharmonic phonons (i.e., $P_\gamma+B_\gamma+B_{\Delta\gamma}$) has a net softening effect 
on the bulk modulus. 
Therefore, beyond achieving a higher computational efficiency, the 
SC-QHA method also provides a convenient and complementary approach to 
the conventional QHA for uncovering the physical origins of thermomechanics. 
\subsection{Algorithm Applications}

\subsubsection{Pressure Dependent Anharmonicity}

\begin{figure}[]
\scalebox{0.55}[0.55]{\includegraphics{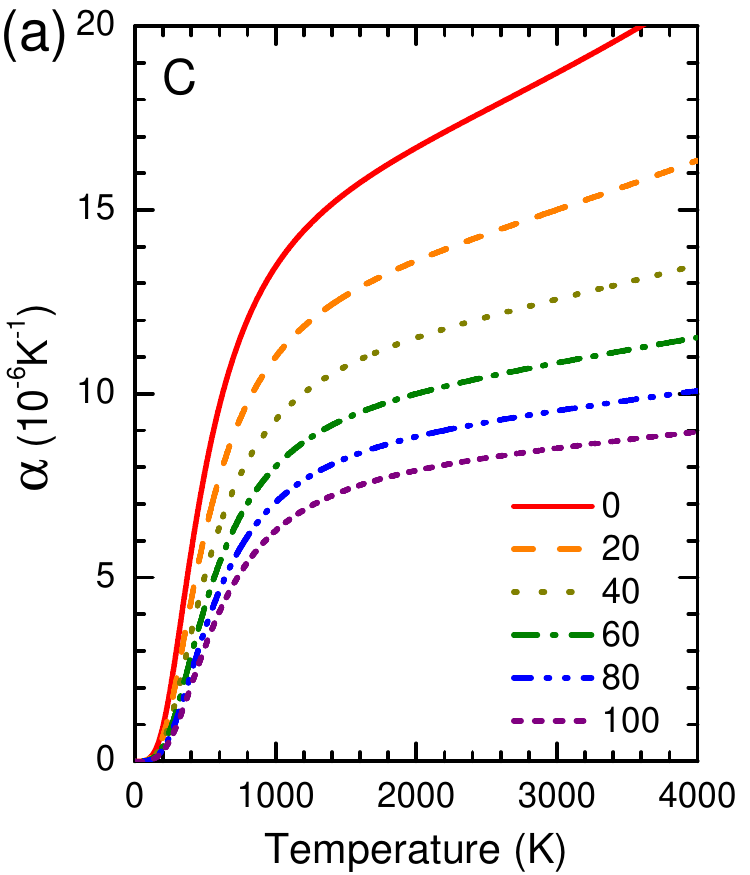}}
\scalebox{0.55}[0.55]{\includegraphics{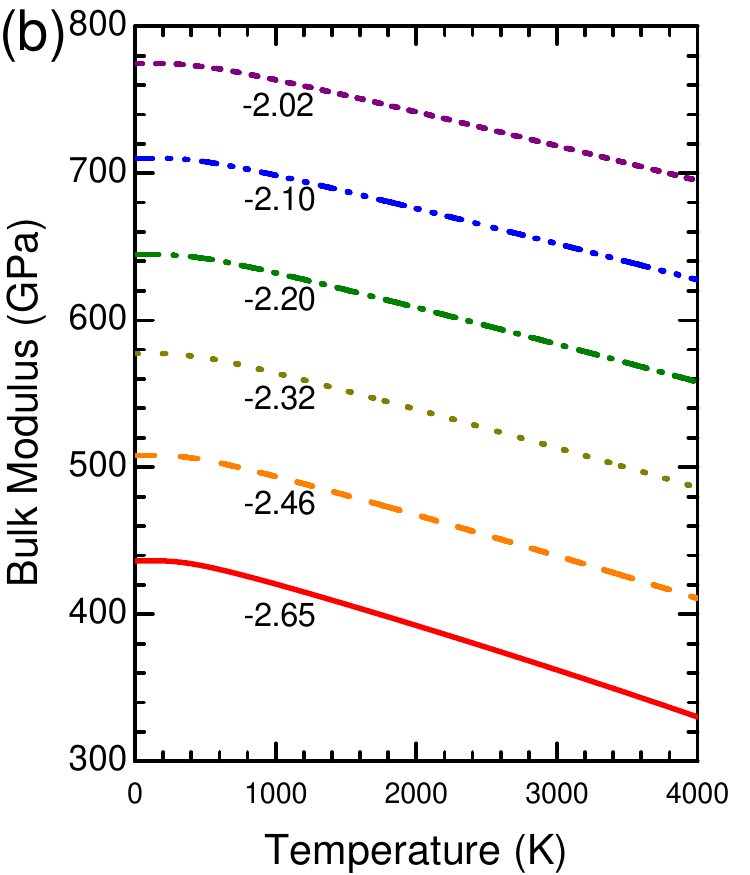}}\\
\scalebox{0.55}[0.55]{\includegraphics{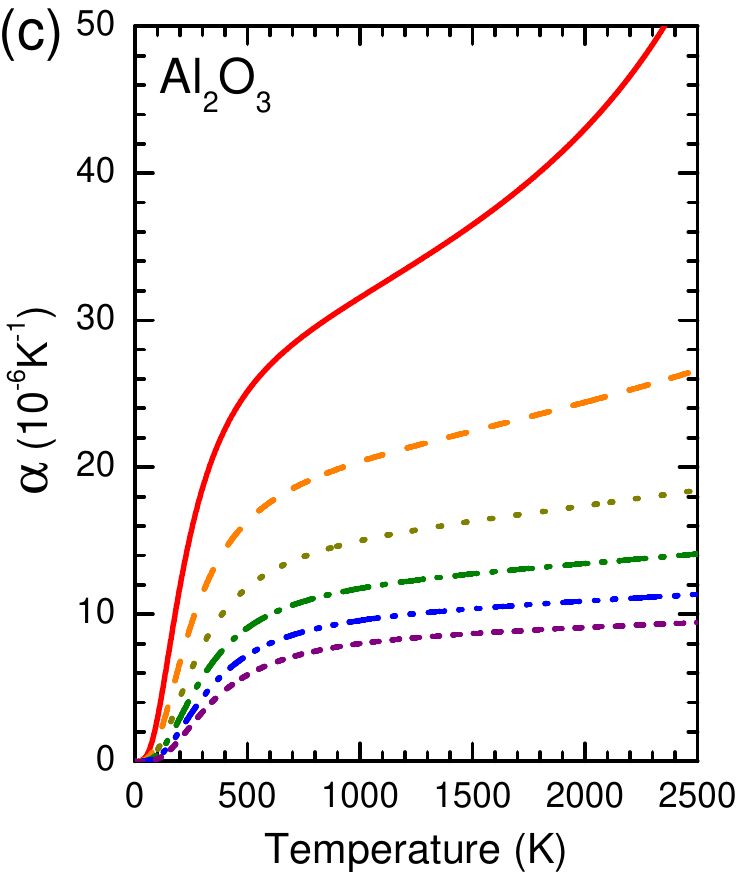}}
\scalebox{0.55}[0.55]{\includegraphics{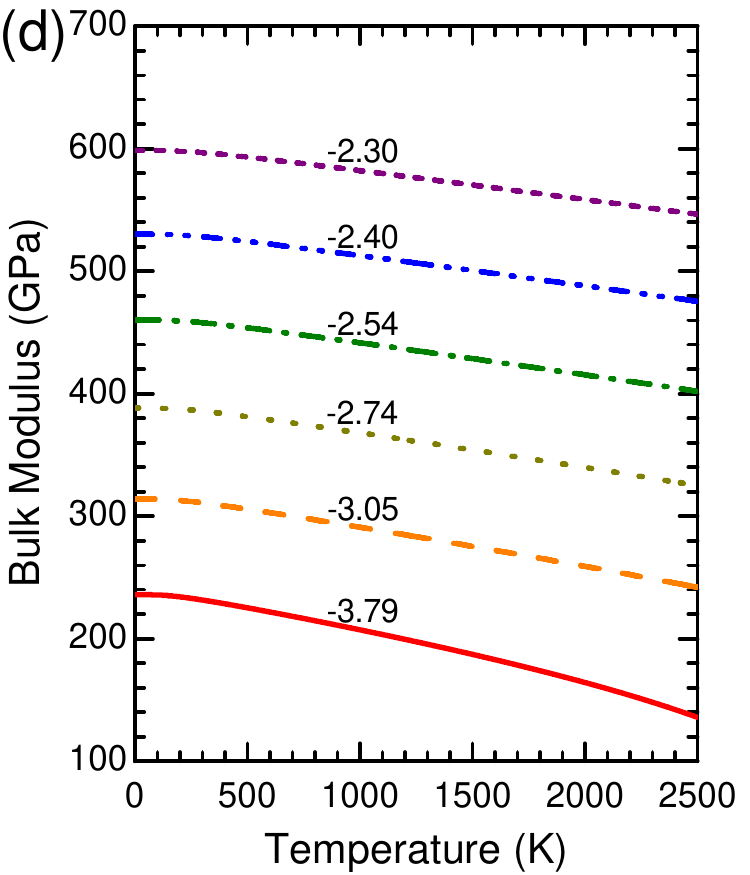}}
\caption{\label{Fig_Pressure_Effect} Pressure dependence of $\alpha$
and $B_T$ for (a,b) diamond and (c,d) alumina, where the slopes (in $10^{-2}$ GPa/K) of the
$B_T$ curves at 1000 K are indicated.
}
\end{figure}

In high-pressure physics, the external pressure is controlled by compressing a diamond anvil
cell (DAC), and the optical spectra of a ruby (Cr-doped Al$_2$O$_3$)
particle adjacent to the sample is used to calculate the actual pressure
in the DAC \cite{Jayaraman1983,Boehler2000,Hemley2001,Mcmahon2015}.
Thus, it is useful to study the pressure dependence of the
anharmonic properties of C and Al$_2$O$_3$. 
When using the conventional QHA method to study the pressure effect, a large range
of volumes should be considered, and the phonon spectra of numerous 
volumes probably need to be calculated to give confidence to the results. 
Considering extra volumes far away from the equilibrium volume, however, 
will introduce some numerical error to the QHA results for 
zero and low pressures, because of the decreased 
weights of the equilibrium and slightly-compressed volumes in the EOS fitting process 
(Supplemental Material). 
Thus, simulating the pressure-dependent
anharmonic properties using the QHA method is not only computationally
expensive, but also prone to intrinsic numerical uncertainties from the 
EOS fitting. 

Here we remedy the shortcomings of the conventional QHA and apply our 
2nd-SC-QHA method to examine the pressure effects on $\alpha$($T$) and $B_T$($T$) for
C and Al$_2$O$_3$ from 0--100 GPa (Fig.\ \ref{Fig_Pressure_Effect}). 
We find that $\alpha$ decreases
with increasing pressure [Fig.\ \ref{Fig_Pressure_Effect}(a,c)], 
which is mainly due to the increased $B_T$ according to Eq. \ref{Equ_alpha} and Fig.\ 
\ref{Fig_Pressure_Effect}(b,d). 
The phonon mode hardening and the corresponding 
decrease in phononic anharmonicity also makes a minor contribution
to the pressure-induced decrease of the thermal expansion coefficient. 
Interestingly, $\alpha$(Al$_2$O$_3$) is approximately two times that of $\alpha$(C) at 0\,GPa, while they 
become comparable at 100\,GPa, because of the faster stiffening rate of $B_T$(Al$_2$O$_3$) with pressure.  
$B_T$ increases with increasing pressure, but decreases with increasing temperature. 
%
The thermal-softening rate given by $-{dB_T}/{dT}$ decreases with increasing pressure
[Fig.\ \ref{Fig_Pressure_Effect}(b,d)], because of the decreased
phonon anharmonicity. When the thermal effects on $\alpha$ and $B_T$ are
comparable with or even larger than the pressure effect, the
simulated results for C and Al$_2$O$_3$ here will be useful for the
high-pressure experiments carried out under variable temperature.
%

\subsubsection{Anharmonicity in Complex Ceramics}

The accuracy of the SC-QHA method for Si, C,
and Al$_2$O$_3$, as well as its application to C and Al$_2$O$_3$
under pressure, motivate us to study the thermal expansion and
thermomechanics of Ca$_3$Ti$_2$O$_7$. 
This compound is an important hybrid-improper 
ferroelectric (HIF) \cite{Benedek2011} that has received considerable attention recently.\cite{10.1038/nmat4168} 
Its thermal expansion has also 
been recently measured in experiment \cite{Senn2015}.

\begin{figure}
\scalebox{0.55}[0.55]{\includegraphics{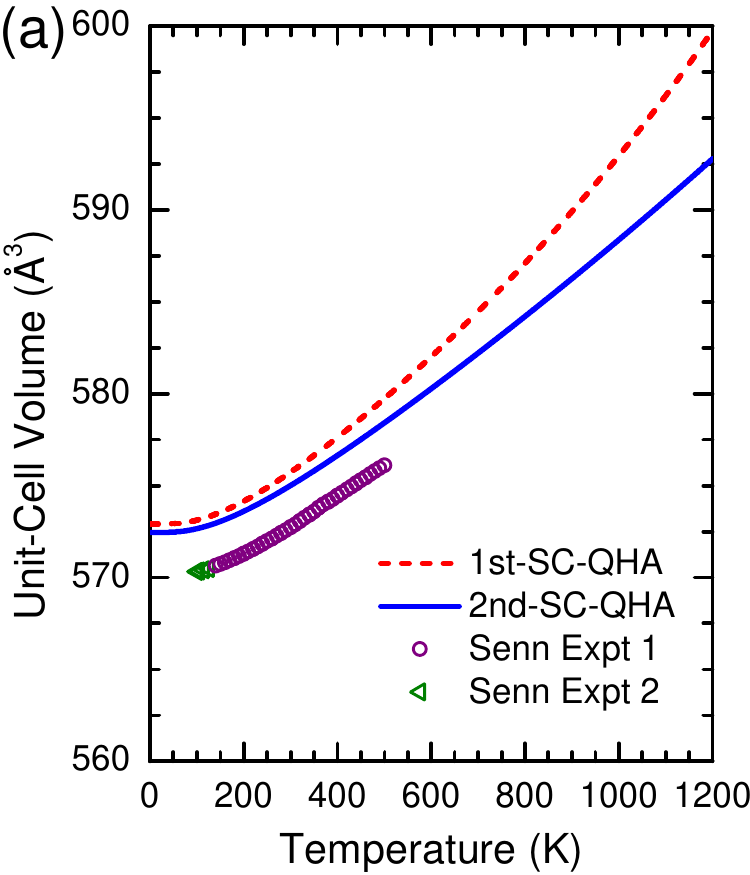}}
\scalebox{0.55}[0.55]{\includegraphics{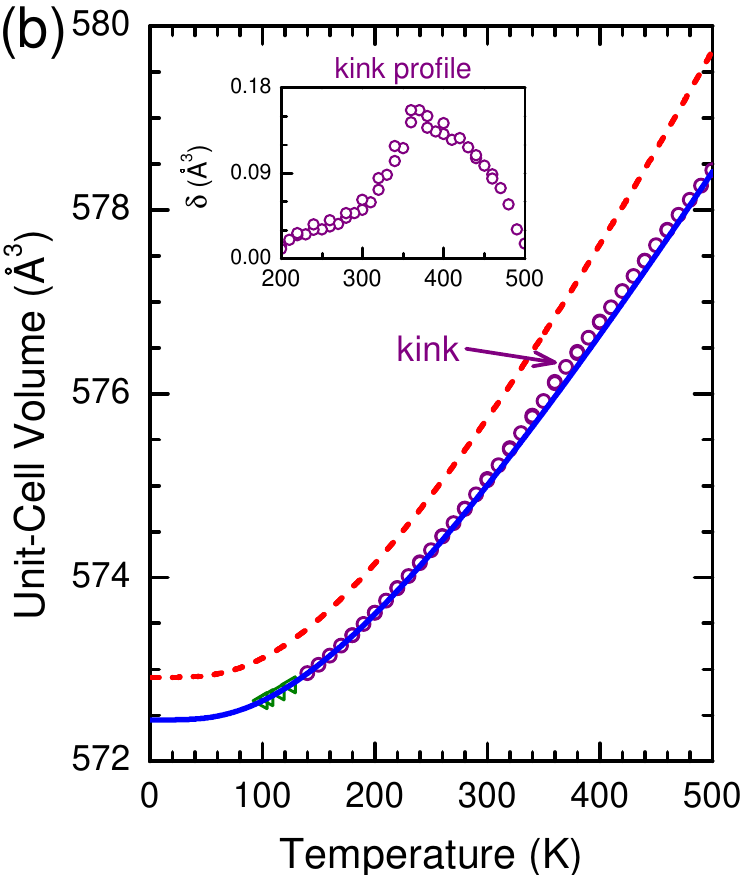}}\\
\scalebox{0.55}[0.55]{\includegraphics{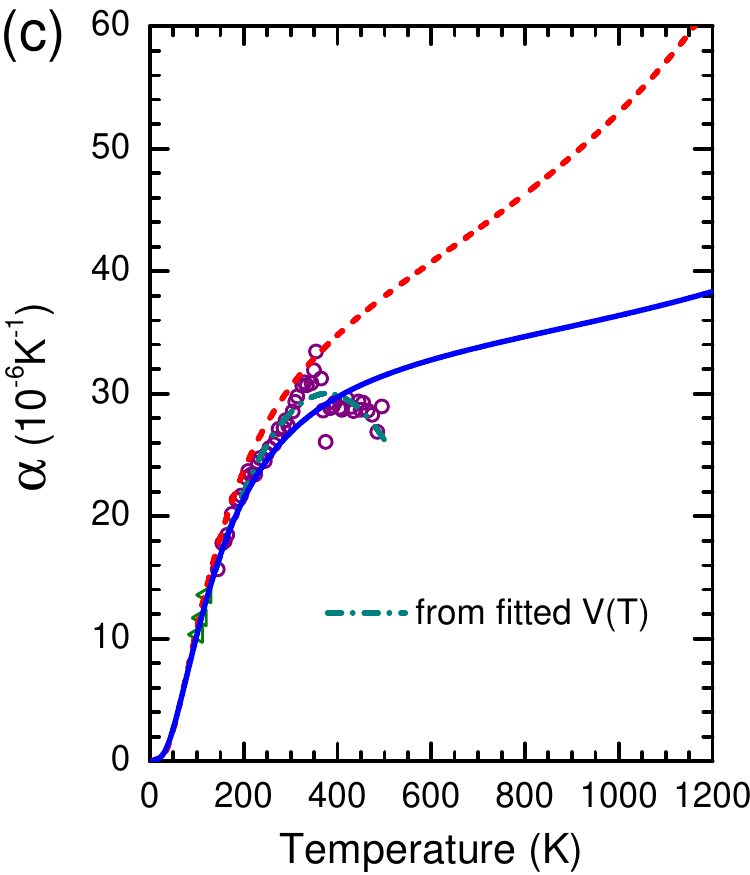}}
\scalebox{0.55}[0.55]{\includegraphics{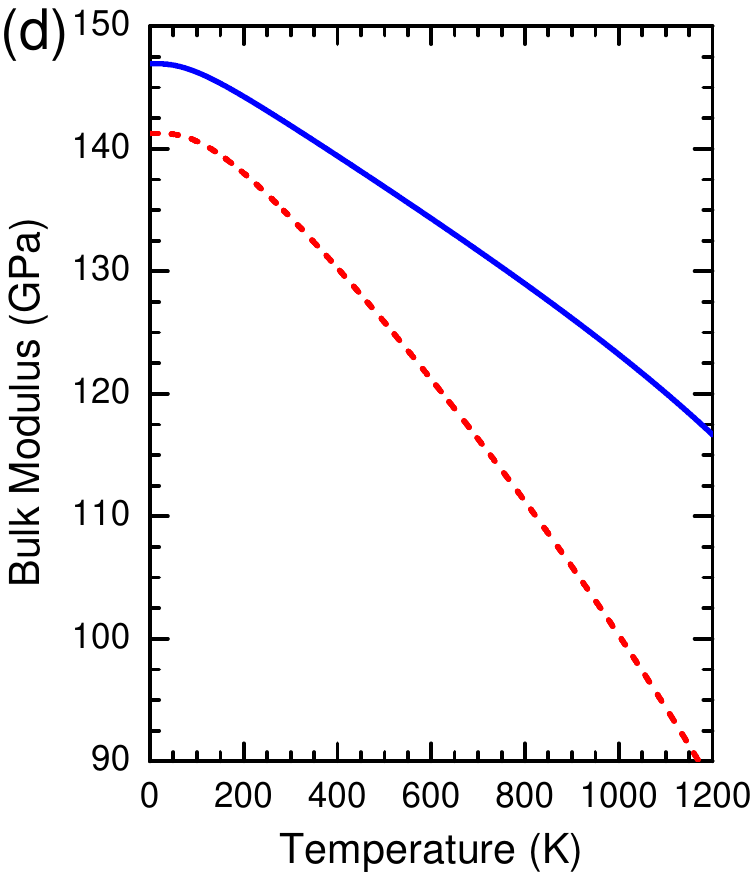}}
\caption{\label{Fig_Ca3Ti2O7} Temperature dependence of the (a,b)
unit-cell volume, (c) thermal expansion coefficient, and (d) bulk
modulus of  Ca$_3$Ti$_2$O$_7$, where the experimental results are shifted in panel (b) to facilitate comparisons.
The first set of experimental results (``Senn Expt 1")
are taken from Ref.~\onlinecite{Senn2015} and the second data set
(``Senn Expt 2") was provided by Senn through a private communication. 
The inset of panel (b) shows a detailed profile of the experimental kink at 360 K, 
where  $\delta$ is the difference between the shifted experimental $V(T)$ data 
and the 2nd-SC-QHA calculate profile.}
\end{figure}

The calculated $V(T)$, $\alpha(T)$, and $B_T(T)$ from both the 1st-SC-QHA and
2nd-SC-QHA methods are shown in \autoref{Fig_Ca3Ti2O7}. 
We find that the 1st-SC-QHA results deviate from the 2nd-SC-QHA ones, 
indicating the importance of the nonlinear frequency-volume variation in this 
ferroelectric material. 
Thus, it is necessary to use the 2nd-SC-QHA method for such kind of HIFs where 
multiple lattice modes interact to stabilize the ferroelectric phase. 
Therefore, we focus our discussion on the results obtained from the 2nd-SC-QHA method below. 

In Fig.\ \ref{Fig_Ca3Ti2O7}(a), our theoretical $V(T)$ variation is
compared with the measured temperature-dependent volume data collected from two 
experiments (labeled as ``Senn Expt 1" and ``Senn Expt 2") by Senn \emph{et al.}
\cite{Senn2015}, one of which (Senn Expt 2) was
provided by Senn and co-workers through private communications after 
publication of Ref.~\onlinecite{Senn2015}. These two sets of experimental data have no 
observable deviation, though according to Senn's
comments, the X-ray wavelengths used for these two sets of
measurements are slightly different, which may introduce a normalization
error in the X-ray diffraction analysis. The 2nd-SC-QHA unit-cell volume is about 0.4\% higher than Senn's 
measurements, indicating the high accuracy of our simulation. 

To facilitate a direct theory-experiment comparison for $V(T)$, the experimental data sets are uniformly 
shifted upwards by 2.3\,\AA$^3$, such that the lowest temperature experimental data coincides 
with our 2nd-SC-QHA $V(T)$ curve as shown in Fig.\ 
\ref{Fig_Ca3Ti2O7}(b). 
We find that our 2nd-SC-QHA $V(T)$ curve perfectly overlaps with the experimental temperature variation,
except for a narrow temperature range, where a very small kink centered at 360\,K appears [Fig.\ \ref{Fig_Ca3Ti2O7}(b), inset].

To have a direct theory-experiment comparison for $\alpha$, we extract the $\alpha$ from the experimental data in 
two ways. 
First, we calculate $\alpha$ from the differential between two neighboring data, i.e., 
\begin{equation}{\label{Equ_expt_alpha_1}}
\alpha\left(\frac{T_1+T_2}{2}\right)=\frac{2}{V_1+V_2}\frac{V_1-V_2}{T_1-T_2}\,.
\end{equation}
Second, we analytically derive $\alpha=\frac{1}{V}\frac{dV}{dT}$  
from a fitted $V(T)$ curve using an exponential function of the form $\sum_{i=0}^3{s_iT^i}$.
All the experimental thermal expansion data agree well with our 2nd-SC-QHA $\alpha(T)$ curve [Fig.\ \ref{Fig_Ca3Ti2O7}(c)],
except for some scattering in experimental data above 300 K caused by the presence of the kink in $V(T)$
[Fig.\ \ref{Fig_Ca3Ti2O7}(b)]. 
This experimental kink may indicate 
some unknown processes, e.g., impurities and domain dynamics, becoming thermally activated 
in the sample above 300 K, or certain small uncontrollable uncertainty in
the experimental characterization. 
Upon decreasing temperature, $\alpha$
decreases down to zero at 0\,K [Fig. \ref{Fig_Ca3Ti2O7}(c)], thus,
the $V(T)$ variation becomes less dispersive [Fig.\ 
\ref{Fig_Ca3Ti2O7}(b)]. 
This is an inevitable quantum-mechanical
effect, where fewer quantized phonons are thermally excited upon
cooling.

Last, from both experimental measurements and our SC-QHA simulation, no negative thermal expansion 
is observed in Ca$_3$Ti$_2$O$_7$. 
%
This has been ascribed by Senn \cite{Senn2015} to the 
octahedra tilting ($X_3^-$ symmetry) being frozen out in the $Cmc2_1$ (or equivalently $A2_1am$ symmetry) phase in the 
Ruddlesden-Popper (RP)  A$_3B_2$O$_7$ (A$=$Ca and $B=$Ti or Mn) compounds.
%
For a more complete understanding of the thermal expansion and other anharmonic properties of Ca$_3$Ti$_2$O$_7$, 
further experimental measurements and more detailed analysis are still required. 
%
%

\section{Conclusions}
A fast and accurate ab-initio method called the  self-consistent
quasiharmonic approximation (SC-QHA) method has been formulated to
calculate various anharmonic properties of solids at finite
temperatures. 
The SC-QHA method not only is about five times per dimension faster
than the conventional QHA method, but also aids in the physical 
analysis of underlying anharmonic mechanisms. 
Although we showed the superior performance of the SC-QHA method compared to the conventional 
QHA using nonmagnetic insulators, the methodology can be readily extended to metallic and magnetic solids, 
where electronic and magnetic excitations are also important. 
Moreover, the basic SC-QHA algorithm can be transferred to the 
realm beyond the quasiharmonic approximation, i.e., to compute 
high-order anharmonicities, and reduce the computational overhead for the simulation of high-order
anharmonic properties.

The efficiency and accuracy calibrations of the SC-QHA
method based on silicon, diamond, and alumina show that the 2nd-order
SC-QHA method is systematically more accurate than the 1st-order
SC-QHA implementation, but the 1st-SC-QHA method is useful for testing 
the computational parameters required in a density functional theory calculation. 
After evaluating the SC-QHA method, we examined the pressure-dependent thermal
expansion and thermomechanics of diamond and alumina, which are two
important materials in high-pressure physics. 
Finally, the SC-QHA method was also used to study the thermal expansion and
thermomechanics of Ca$_3$Ti$_2$O$_7$, which is structurally complex
and computationally challenging for the  conventional QHA method. 
No negative thermal expansion was found in Ca$_3$Ti$_2$O$_7$, 
which is consistent with a recent experimental measurement. 
The simulated $V(T)$ variation also precisely agrees with available experimental data. 
These results demonstrate that the SC-QHA method is
both an efficient computational and useful theoretical tool to understand experimentally 
determined anharmonic properties of materials.

\begin{acknowledgments}
L.-F.H.\ and J.M.R.\ wish to thank Dr.\ M.\ Senn (University of
Oxford), and Prof.\ S.-W.\ Cheong (Rutgers University) for
providing additional experimental data for
Ca$_3$Ti$_2$O$_7$, as well as for the helpful e-mail exchanges.
L.-F.H.\ and E.T.\ were supported by the Office of Naval Research MURI
``Understanding Atomic Scale Structure in Four Dimensions to Design
and Control Corrosion Resistant Alloys'' under Grant No.\
N00014-14-1-0675. X.-Z.L.\ and J.M.R.\ were supported by the National Science Foundation
(NSF) through the Pennsylvania State University MRSEC under
award number DMR-1420620. 
Calculations were performed using the QUEST HPC
Facility at Northwestern University and the HPCMP facilities at the
Navy DoD Supercomputing Resource center.\\
\end{acknowledgments}

\section*{Contributions}
The study was planned, methods formulated, calculations carried out, and the manuscript
prepared by L.-F.H.\ and J.M.R. 
X.-Z.L.\ and E.T.\  performed analyses on Ca$_3$Ti$_2$O$_7$ and 
Al$_2$O$_3$, respectively.
All authors discussed the results, wrote, and commented on the
manuscript.
\bibliography{Reference_list}

\clearpage

\end{document}